\PassOptionsToPackage{table}{xcolor} %
\documentclass[sigconf]{acmart}
\pdfoutput=1
\usepackage{hyperref}
\usepackage[table]{xcolor}
\usepackage{tabularx}
\usepackage{booktabs}
\usepackage{multirow}
\usepackage{array}

\usepackage{amsmath}
\usepackage{amsfonts}
\usepackage{nicefrac}
\usepackage{cases}
\usepackage{bm}

\usepackage[utf8]{inputenc}
\usepackage{pifont}
\usepackage{xspace}
\usepackage{hyphenat} %
\usepackage{balance}
\usepackage{enumitem} %
\usepackage{soul} %

\usepackage{algorithm}
\usepackage[noend]{algpseudocode}

\usepackage{tikz}
\usetikzlibrary{fit,positioning,calc,shapes,backgrounds}
\usepackage{pgfplots}
\pgfplotsset{compat=1.14}
\usepgfplotslibrary{fillbetween, groupplots}
\usepackage{caption}
\usepackage{subcaption}

\newcommand{\mpara}[1]{\medskip\noindent{\bf #1}}

\definecolor{cycle1}{RGB}{235,172,35}
\definecolor{cycle2}{RGB}{184,0,88}
\definecolor{cycle3}{RGB}{0,140,249}
\definecolor{cycle4}{RGB}{0,110,0}
\definecolor{cycle5}{RGB}{0,187,173}
\definecolor{cycle6}{RGB}{209,99,230}
\definecolor{cycle7}{RGB}{178,69,2}
\definecolor{cycle8}{RGB}{255,146,135}
\definecolor{cycle9}{RGB}{89,84,214}
\definecolor{cycle10}{RGB}{0,198,248}
\definecolor{cycle11}{RGB}{135,133,0}
\definecolor{cycle12}{RGB}{0,167,108}
\definecolor{cyclegray}{RGB}{189,189,189}

\newtheorem{definition}{Definition}

\newcommand{\win}{\cellcolor{cycle4!25}}

\newcommand{\thiswork}{\textsc{SLaQ}\xspace}

\newcommand*{\bigO}{\mathcal{O}}

\DeclareMathOperator{\tr}{tr}

\newcommand*{\vH}{\mathbf{h}}
\newcommand*{\vV}{\mathbf{v}}
\newcommand*{\vU}{\mathbf{u}}
\newcommand*{\vQ}{\mathbf{q}}
\newcommand*{\vE}{\mathbf{e}}

\newcommand*{\mA}{\mathbf{A}}
\newcommand*{\mD}{\mathbf{D}}
\newcommand*{\mI}{\mathbf{I}}
\newcommand*{\mH}{\mathbf{H}}
\newcommand*{\mHe}{\bm{\mathcal{H}}}
\newcommand*{\mL}{\mathbf{L}}
\newcommand*{\mP}{\mathbf{P}}
\newcommand*{\mLn}{\bm{\mathcal{L}}}
\newcommand*{\mPhi}{\bm{\Phi}}
\newcommand*{\mLambda}{\bm{\Lambda}}

\newcommand*{\mQ}{\mathbf{Q}}
\newcommand*{\mT}{\mathbf{T}}

\newcommand*{\mU}{\mathbf{U}}

\newcommand*{\sR}{\mathbb{R}}
\newcommand*{\sN}{\mathbb{N}}

\newcommand{\ddblp}{\textsc{DBLP}\xspace}

\newcommand{\dwikim}[1]{\textsc{Wiki-#1}\xspace}
\newcommand{\dorkut}{\textsc{SNAP-Orkut}\xspace}
\newcommand{\dfriendster}{\textsc{Friendster}\xspace}
\newcommand{\dclueweb}{\textsc{ClueWeb09}\xspace}
\newcommand{\dlivejournal}{\textsc{LiveJournal}\xspace}
\newcommand{\ddd}{\textsc{D\&D}\xspace}
\newcommand{\dcollab}{\textsc{COLLAB}\xspace}
\newcommand{\dredditm}{\textsc{Reddit-5k}\xspace}
\newcommand{\dredditmb}{\textsc{Reddit-12k}\xspace}
\title{Just SLaQ When You Approximate: Accurate~Spectral~Distances~for~Web-Scale~Graphs}

\copyrightyear{2020}
\acmYear{2020}
\setcopyright{iw3c2w3}
\acmConference[WWW '20]{Proceedings of The Web Conference 2020}{April 20--24, 2020}{Taipei, Taiwan}
\acmBooktitle{Proceedings of The Web Conference 2020 (WWW '20), April 20--24, 2020, Taipei, Taiwan}
\acmPrice{}
\acmDOI{10.1145/3366423.3380026}
\acmISBN{978-1-4503-7023-3/20/04}
 
\begin{document}
\author{Anton Tsitsulin}
\authornote{Equal contribution.}
\authornote{Work done while interning at Google.}
\affiliation{%
  \institution{University of Bonn}
}
\email{tsitsulin@bit.uni-bonn.de}
\author{Marina Munkhoeva}
\authornotemark[1]
\authornote{Work done while interning at X, the Moonshot Factory, formerly Google X.}
\affiliation{%
  \institution{Skoltech}
}
\email{marina.munkhoeva@skoltech.ru}
\author{Bryan Perozzi}
\affiliation{%
  \institution{Google Research}
}
\email{bperozzi@acm.org} %

\begin{abstract}

Graph comparison is a fundamental operation in data mining and information retrieval.
Due to the combinatorial nature of graphs, it is hard to balance the expressiveness of the similarity measure and its scalability.
Spectral analysis provides quintessential tools for studying the multi-scale structure of graphs and
is a well-suited foundation for reasoning about differences between graphs.
However, computing full spectrum of large graphs is computationally prohibitive; thus, spectral graph comparison methods
often rely on rough approximation techniques with weak error guarantees.

In this work, we propose \thiswork{}, an efficient and effective approximation technique for computing spectral distances between graphs with billions of nodes and edges.
We derive the corresponding error bounds and demonstrate that accurate computation is possible in time linear in the number of graph edges.
In a thorough experimental evaluation, we show that \thiswork outperforms existing methods, oftentimes by several orders of magnitude in approximation accuracy, and maintains comparable performance, allowing to compare million-scale graphs in a matter of minutes on a single machine.

\end{abstract} \maketitle
\section{Introduction}\label{sec:introduction}

Many complex systems, including social and biological, and interactions on the Web can be concisely modeled as graphs.
Solving data mining tasks such as classification or anomaly detection on graphs requires sophisticated techniques.
However, if one has a meaningful notion of similarity between two graphs, classic data mining techniques can be effortlessly applied to graphs.
As many real-world graphs are huge (millions of nodes and edges), recent research~\cite{koutra2013,tsitsulin2018kdd,chen2019} has focused on providing \emph{scalable} graph distances.

\begin{figure}[!t]
\centering
\subcaptionbox{Dolphin and karate graphs}[.5\columnwidth]{
\centering
\resizebox{.5\columnwidth}{!}{%
\begin{tikzpicture}
\foreach \nodename/\x/\y in {9/0.09982232/-0.25192524, 4/-0.09943474000000001/-0.37490623, 10/3.703901/0.5120902, 6/3.8205164000000003/0.29763186, 7/3.4907672000000005/0.5540524, 11/-0.1488507/0.5032162, 1/0.029728150000000002/0.22932562, 3/-0.41119132999999997/0.8719469, 14/3.5510254/0.39347740000000003, 15/-0.7897327000000001/0.0034600666, 16/-0.20206137/-0.7133409, 17/-0.731151/0.25390356000000003, 18/3.1919348000000003/0.7433347, 2/2.2228784/0.46718722999999995, 19/-0.6394937500000001/-0.79086075, 20/1.9499261/0.49293433999999997, 8/1.6723549/0.44424416, 21/-0.15116543000000002/0.15413552, 22/-0.8485212/-0.6101515, 23/3.3115744/1.0238649, 25/-0.7931284/-0.8534014999999999, 26/2.7885809999999998/0.9170456, 27/2.5275578/0.8601411400000001, 28/2.4924239/0.7326126, 29/0.8312946/0.31229364, 30/-1.0200581/-0.5746090300000001, 31/1.08368675/0.53990215, 32/3.4073877/0.9896026600000001, 33/4.105993000000001/0.5203302, 34/-1.0633990500000001/0.064443274, 13/-1.5159605/0.018357946, 35/-1.1992307/0.43529804, 36/-1.3877332999999998/-0.78892624, 37/0.641737/-0.17896435, 24/-0.033058825/-0.90715935, 38/-0.57394566/0.063031435, 39/-0.9847513999999999/0.3992852, 40/2.0311087/-0.056099977, 41/-0.04801807/-0.04672297, 42/3.1203879999999997/0.37888214, 43/0.14349835/0.57580803, 44/-1.3441782/0.20464937, 45/-0.72882385/0.67814766, 46/-0.43690154999999997/-0.7191442000000001, 47/-1.8355873/0.4546721, 48/0.4109523/0.49232210000000004, 50/-1.7639752/0.58355503, 51/-0.52317554/-0.19232613, 52/-0.7026662/-1.1242351, 5/-0.879844/-1.4880661000000002, 12/-0.7813308000000001/-1.5205840000000002, 53/-0.9003067/-0.16899342, 54/-1.3628301999999999/0.7361864, 55/2.7784738/0.39534058000000005, 56/-0.39923977/-1.2322289000000002, 57/3.9284097000000004/0.42525084999999996, 58/3.3006708/0.25832905, 49/3.487004/0.0041569182, 59/-1.2273173/0.7340483999999999, 60/0.15178095/-0.5826551400000001, 61/4.400272800000001/0.5548184, 62/-0.8923716/0.82084076}
{\node (n\nodename) at (\x,\y) [shape=circle,inner sep=2pt,draw,thick,fill=cycle1] {};}
\begin{pgfonlayer}{background}
\path
\foreach \startnode/\endnode in {n9/n4, n10/n6, n10/n7, n11/n1, n11/n3, n14/n6, n14/n7, n14/n10, n15/n1, n15/n4, n16/n1, n17/n15, n18/n2, n18/n7, n18/n10, n18/n14, n19/n16, n20/n2, n20/n8, n21/n9, n21/n17, n21/n19, n22/n19, n23/n18, n25/n15, n25/n16, n25/n19, n26/n18, n27/n2, n27/n26, n28/n2, n28/n8, n28/n18, n28/n26, n28/n27, n29/n2, n29/n9, n29/n21, n30/n11, n30/n19, n30/n22, n30/n25, n31/n8, n31/n20, n31/n29, n32/n18, n33/n10, n33/n14, n34/n13, n34/n15, n34/n17, n34/n22, n35/n15, n35/n34, n36/n30, n37/n2, n37/n21, n37/n24, n38/n9, n38/n15, n38/n17, n38/n22, n38/n34, n38/n35, n38/n37, n39/n15, n39/n17, n39/n21, n39/n34, n40/n37, n41/n1, n41/n8, n41/n15, n41/n16, n41/n34, n41/n37, n41/n38, n42/n2, n42/n10, n42/n14, n43/n1, n43/n3, n43/n11, n43/n31, n44/n15, n44/n30, n44/n34, n44/n38, n44/n39, n45/n3, n45/n21, n45/n35, n45/n39, n46/n9, n46/n16, n46/n19, n46/n22, n46/n24, n46/n25, n46/n30, n46/n38, n47/n44, n48/n1, n48/n11, n48/n21, n48/n29, n48/n31, n48/n43, n50/n35, n50/n47, n51/n15, n51/n17, n51/n21, n51/n34, n51/n43, n51/n46, n52/n5, n52/n12, n52/n19, n52/n22, n52/n24, n52/n25, n52/n30, n52/n46, n52/n51, n53/n15, n53/n30, n53/n39, n53/n41, n54/n44, n55/n2, n55/n7, n55/n8, n55/n14, n55/n20, n55/n42, n56/n16, n56/n52, n57/n6, n57/n7, n58/n6, n58/n7, n58/n10, n58/n14, n58/n18, n58/n40, n58/n42, n58/n49, n58/n55, n59/n39, n60/n4, n60/n9, n60/n16, n60/n37, n60/n46, n61/n33, n62/n3, n62/n38, n62/n54}
{
(\startnode) edge[-,thick,draw opacity=0.75] node {} (\endnode)
};
\end{pgfonlayer}
\end{tikzpicture} %
}
\resizebox{.33\columnwidth}{!}{%
\begin{tikzpicture}
\foreach \nodename/\x/\y/\nc in {
0/-1.1620834335566734/1.6657525642583222/94.117647058821348, 1/-1.1892584131033608/1.3700792964387705/52.941176470586917, 2/-2.10241580598485/0.8160695579824446/58.823529411763964, 3/-0.7021651436094368/1.0207708374817235/35.294117647057845, 4/-0.796934961950899/2.5662586824091953/17.647058823528859, 5/-1.163808781445275/2.7342972855064698/23.529411764705248, 6/-1.524512792595378/2.8461770009569385/23.529411764705273, 7/-0.6815254310138741/0.6655088412109437/23.529411764705316, 8/-2.7039841430278/1.4033339988272575/29.411764705882192, 9/-2.3843397007704388/0.067743386762977/11.764705882352926, 10/-0.493408147964988/2.364606219391841/17.647058823528802, 11/-0.12279791486376579/0.9829755609740104/5.882352941176279, 12/-0.0042696362140608474/1.6124692664139062/11.764705882352592, 13/-1.7833448687734768/1.116572640499568/29.411764705881826, 14/-4.763939073345353/0.8154751614117168/11.764705882352994, 15/-4.934960560131771/1.8127504243311172/11.764705882352997, 16/-2.2562081451933547/2.9592999991305717/11.764705882352692, 17/-0.13166507200014152/1.9564770797314988/11.764705882352610, 18/-5.0/1.5709858460870798/11.764705882352994, 19/-2.1919166775251124/1.839068186312868/17.647058823529100, 20/-4.918155041217618/1.0643738754075218/11.764705882352990, 21/-0.0/1.3430321761449622/11.764705882352612, 22/-4.3483573637984305/2.198106871176102/11.764705882352994, 23/-4.5110705317224005/0.5740158106087705/29.411764705882476, 24/-3.0940611735440426/0.0/17.647058823529374, 25/-3.7379532234315422/0.15334470455962784/17.647058823529381, 26/-4.80902754655218/2.046635622610528/11.764705882353001, 27/-3.6592828415811147/0.3770857983410918/23.529411764705902, 28/-2.837615171301133/0.3191331539747288/17.647058823529310, 29/-4.994646116375242/1.3122509016595434/23.529411764706033, 30/-2.982097711274267/1.7449840810187065/23.529411764705639, 31/-2.9112945716970735/0.7720185968546954/35.294117647058677, 32/-4.024548804484681/1.3204243037025896/70.588235294117808, 33/-3.6963720111992493/1.1806803216661164/100.000000000000000
}
{
\node (\nodename) at (\x,\y) [shape=circle,inner sep=3pt,draw,thick,fill=cycle3] {}; %
}
\begin{pgfonlayer}{background}
\path
\foreach \startnode/\endnode in {
0/1, 0/2, 0/3, 0/4, 0/5, 0/6, 0/7, 0/8, 0/10, 0/11, 0/12, 0/13, 0/17, 0/19, 0/21, 0/31, 1/17, 1/2, 1/3, 1/21, 1/19, 1/7, 1/13, 1/30, 2/3, 2/32, 2/7, 2/8, 2/9, 2/27, 2/28, 2/13, 3/7, 3/12, 3/13, 4/10, 4/6, 5/16, 5/10, 5/6, 6/16, 8/32, 8/30, 8/33, 9/33, 13/33, 14/32, 14/33, 15/32, 15/33, 18/32, 18/33, 19/33, 20/32, 20/33, 22/32, 22/33, 23/32, 23/25, 23/27, 23/29, 23/33, 24/25, 24/27, 24/31, 25/31, 26/33, 26/29, 27/33, 28/33, 28/31, 29/32, 29/33, 30/33, 30/32, 31/32, 31/33, 32/33}
{
(\startnode) edge[-,thick,draw opacity=0.75] node {} (\endnode)
};
\end{pgfonlayer}
\end{tikzpicture} %
}
}
\subcaptionbox{Spectral descriptors}[.4\columnwidth]{
\centering
\resizebox{.43\columnwidth}{!}{%
\begin{tikzpicture}
	\begin{axis}[
		ylabel={NetLSD},
		xlabel={$t$},
		xmin=0.01,
		xmax=100,
		ymin=0,
		ymax=1,
		xmode=log,
		ymode=log,
		width=0.25\textwidth,
		height=3.5cm,
	]
	\addplot[very thick,color=cycle1] table[x=t,y=lsd] {data/karate-lsd.dat};
	\addplot[very thick,color=cycle3] table[x=t,y=lsd] {data/dolphins-lsd.dat};
	\end{axis}%
\node[inner sep=0pt] at (1.5,-1.15) {Dolphins~$\mHe=3.846$};
\node[inner sep=0pt] at (1.67,-1.55) {Karate~$\mHe=3.154$};
\end{tikzpicture} %
}
}
\caption{Spectral analysis provides valuable insights in the structure of graphs. Can we scale it to billions of nodes?}
\label{fig:my_label}
\end{figure}
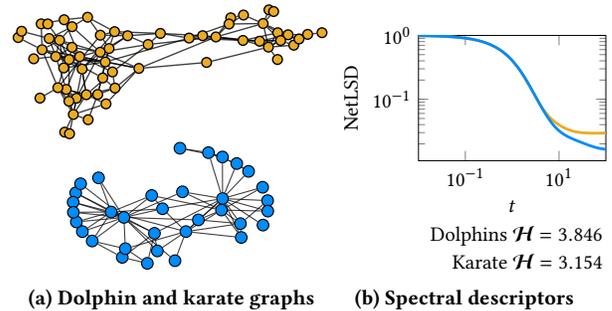
 
Spectral analysis provides powerful methods for graph clustering~\cite{vonluxburg2007,buhler2009,nascimento2011}, comparison~\cite{tsitsulin2018kdd,chen2019}, alignment~\cite{nassar2018,heimann2018}, and characterization of graphs~\cite{estrada2000,han2012,gutman2006,tsitsulin2019www,chen2019}.
In practice, however, the applicability of these methods is often limited by the scalability of eigendecomposition itself: it takes cubic time to compute all eigenvalues and eigenvectors of a given graph.
Several graph comparison methods such as Von Neumann Graph Entropy (VNGE)~\cite{braunstein2006,chen2019} and NetLSD~\cite{tsitsulin2018kdd} require only information derived from a graph's eigenvalues -- not the full decomposition.
Although they depend on less information, na\"ive computation of such metrics is just as expensive as a full eigendecomposition.
To make these methods scale to large graphs, these methods resort to low-order (two terms) Taylor expansion having loose bounds and poor empirical approximation performance.
Few works discuss approximation accuracy or experiment on how it affects performance on downstream tasks.

In this work, we propose \thiswork{}, an approximation algorithm for computing spectral distances for very large graphs.
By leveraging recent advances in numerical linear algebra~\cite{ubaru2017,adams2018,cohensteiner2018,dong2019}, we achieve state-of-the-art approximation accuracy in time linear in the number of graphs' edges.

We summarize our contributions:
\begin{itemize}[leftmargin=0.2cm,itemindent=.4cm,labelwidth=\itemindent,labelsep=0cm,align=left]
\item We introduce \thiswork, an efficient approximation technique for two spectral graph distances, VNGE and NetLSD.
\item We derive corresponding approximation error bounds and experimentally observe an average reduction in the approximation error of $\mathbf{30\!\times\!-200\times}$ over a diverse set of real-world graphs.
\item We demonstrate that faithful approximation is \emph{necessary} for accurate graph comparison and current approximation techniques are unfit for accurate yet fast approximation.
\item We show that accurate computation of VNGE and NetLSD is possible for a graph with billions of nodes and edges on a single machine \emph{in less than an hour}.
\end{itemize}

\section{Preliminaries}\label{sec:background}

We introduce two techniques, NetLSD~\cite{tsitsulin2018kdd} and VNGE~\cite{braunstein2006,chen2019}, and establish basic notation and facts about these techniques that we will need for accurate approximation.
Table~\ref{tab:symbols} summarizes the notation used throughout the paper

Let $G=(V, E)$ be an undirected graph,  represented as a set of vertices $V = (v_1, \ldots, v_n), |V| = n$ and a set of edges $E \subseteq V\times V, |E| = m$.
The adjacency matrix $\mA$ is a $n \times n$ matrix having a positive weight ($1$ if the graph is unweighted) $\mA_{ij} > 0$ associated with each edge $(v_i, v_j)$ and $0$ otherwise.
There are several matrices important for spectral analysis that are associated with that graph.
We consider two: the \emph{Laplacian} matrix $\mL = \mD - \mA$ and the \emph{normalized Laplacian} matrix $\mLn = \mI - \mD^{\nicefrac{-1}{2}}\mA \mD^{\nicefrac{-1}{2}}$, where $\mD$ is the diagonal matrix with the degree of the node $i$ as entry $\mD_{ii}$, i.e.\ $\mD_{ii} = \sum_{j = 1}^n \mA_{ij}$ and $\mI$ is the identity matrix.

As both Laplacians are symmetric, they can be factorized as $\mL = \mPhi\mLambda\mPhi^\top$, where $\mLambda$ is a diagonal matrix on the sorted eigenvalues $\lambda_1 \le \ldots \le \lambda_n$ of which $\phi_1, \ldots, \phi_n$ are the corresponding eigenvectors, and $\mPhi$ is an orthogonal matrix obtained by stacking the eigenvectors in columns $\mPhi = [\phi_1 \phi_2 \ldots \phi_n]$.
The set of eigenvalues $\{\lambda_1, \ldots, \lambda_n\}$ is called the \emph{spectrum} of a graph.
We note that although NetLSD and VNGE operate on different Laplacian matrices, for convenience's sake we will refer to the spectrum of both matrices as $\lambda_i$.

Spectrum of the Normalized Laplacian $\mLn$ is bounded to $[0; 2]$, for the unnormalized counterpart it is $[0; 2\max_i \mD_{ii}]$ as per the Gershgorin circle theorem~\cite{gershgorin1931,golub2009}.
The two Laplacians each reflect different properties of a graph.
For example, the normalized Laplacian can not distinguish number of edges~\cite{chung1997} (as it operates on local densities), however its second eigenvector can be used to optimize normalized cut size~\cite{shi2000}.

\begin{table}[!t]
\begin{center}
\small
\setlength{\tabcolsep}{1pt}
\renewcommand{\aboverulesep}{0.3pt}
\renewcommand{\belowrulesep}{0.3pt}
\newcolumntype{C}{>{\centering\arraybackslash}X}
\begin{tabularx}{\columnwidth}{p{2.5cm}X}
\textbf{Symbol} & \textbf{Description} \\ 
\midrule
$n\in\sN^+$ & $|V|=$ number of vertices \\
$m\in\sN^+$ & $|E|=$ number of vertices \\
$i,j\in\sN^+$ & vertex indices \\
\midrule
$\mA\in\sR^{n\times n}$ & Adjacency matrix of a graph \\
$\mD\in\sR^{n\times n}$ & Diagonal degree matrix $\mD_{ii} = \sum_{j = 1}^n \mA_{ij}$ \\
$\mL\in\sR^{n\times n}$ & Graph Laplacian $\mL = \mD - \mA$ \\
$\mLn\in\sR^{n\times n}$ & Normalized Laplacian matrix $\mLn = \mI - \mD^{\nicefrac{-1}{2}}\mA \mD^{\nicefrac{-1}{2}}$ \\
\midrule
$\phi\in\sR^{n}; \mPhi\in\sR^{n\times n}$ & Eigenvector; stacked eigenvector matrix \\
$\lambda\in\sR; \mLambda\in\sR^{n\times n}$ & Eigenvalue; diagonal eigenvalue matrix \\
\midrule
$n_v\in\sR^+$ & Number of SLQ random vectors \\
$s\in\sR^+$ & Number of SLQ Lanczos iterations \\
$\vV\in\sR^{n}$ & Random normal vector \\
$f(\mL)\in\sR^{n\times n}$ & Matrix function of graph Laplacian \\
\midrule
$\mHe\in\sR^+$ & VNGE \\
$\vH\in\sR^d$ & NetLSD \\
\bottomrule
\end{tabularx}
\end{center}
\caption{Summary of notation used throughout the paper.}\label{tab:symbols}
\end{table} 
\subsection{Von Neumann Graph Entropy}\label{ssec:vnge}
In the Standard Quantum Mechanics model, the state of a quantum mechanical system associated with the $n$-dimensional Hilbert space is identified with a $n\times n$ positive semidefinite, trace-one, Hermitian \emph{density matrix}.
Von Neumann entropy~\cite{vonneumann2013} is a quantitative measure of mixedness of this density matrix, and is defined as follows:
\begin{definition}
\textbf{Von Neumann Graph Entropy} $\mHe$ is defined as $\mHe = -\sum_i\lambda_i\ln\lambda_i$. VNGE is completely determined by the spectrum.
\end{definition}

\noindent By convention, $0\log0=0$. Braunstein~et~al.~\cite{braunstein2006} reinterprets the graph Laplacian matrix $\mL$ as a quantum mechanical system and introduces Von Neumann Graph Entropy (VNGE) by scaling graph Laplacian $\mL$ by its trace to get the density matrix $\mP = \frac{1}{\tr(\mL)} \mL$.
Scaling the Laplacian does not affect the shape of its spectrum, as each eigenvalue is simply multiplied by $\frac{1}{\tr(\mL)}$.

VNGE is related to the centralization of graphs~\cite{simmons2018}, however, general structural interpretation of this measure is unknown mainly due to the lack of accurate scalable approximation~\cite{han2012,minello2018}.

\subsection{Network Laplacian Spectral Descriptors}\label{ssec:netlsd}

Tsitsulin~et~al.~\cite{tsitsulin2018kdd} introduced NetLSD as a spectral distance between graphs grounded in differential geometry.
Instead of directly operating on the Laplacian matrix, NetLSD is defined in terms of the \emph{heat kernel} of a graph.
The heat equation associated with the Laplacian is:
\begin{equation}\label{eq:heat-diffeq}
    \frac{\partial \vU_t}{\partial t} = -\mLn\vU_t,
\end{equation}
\noindent where $\vU_t\in\sR^{n}$ is a vector representing the heat of each vertex at time $t$.
The solution to the heat equation provides the heat at each vertex at time $t$, when the initial heat $\vU_0$ is initialized with the same value on all vertices.
Its closed-form solution is given by the \emph{heat kernel} matrix $\mH_t\in\sR^{n\times n}$ that can be computed by directly exponentiating the Laplacian in the spectral domain~\cite{chung1997}:
\begin{equation}\label{eq:heat-solution}
    \mH_t = e^{-t\mLn} = \mPhi e^{-t\mLambda}\mPhi^\top = \sum_{i=1}^n e^{-t \lambda_i} \phi_i \phi_i^\top,
\end{equation}
\noindent where ${\mH}_{ij}$ represents the amount of heat transferred from vertex $v_i$ to vertex $v_j$ at time $t$.

The \emph{trace} of the heat kernel matrix provides a useful lower bound on the Gromov-Wasserstein distance between the underlying manifolds~\cite{memoli2011}:
\begin{equation}\label{eq:heat-trace}
    \vH_t = \tr(\mH) = \sum_{i}{e^{-t\lambda_i}}.
\end{equation}
\noindent Then the NetLSD representation is a \textbf{heat trace signature} of graph $G$, i.e., a collection of heat traces at different \emph{time scales}, $\vH(G) = {\{ \vH_t \}}_{t>0}$.
In practice, $t$ is sampled on a logarithmically spaced grid, so $\vH(G)\in\sR^{d}$ is a vector of some small fixed dimensionality $d$.

\subsection{Approximation methods}\label{sec:approximation}
Both VNGE and NetLSD can be represented as a function $\tr f(\mLambda)$ of Laplacian eigenvalues.
A na{\"i}ve approach would be to compute the exact eigenvalues and compute that function as $\sum_i f(\lambda_i)$, however, as we mentioned before, the computational complexity of full eigendecomposition is $\bigO(n^3)$, which is infeasible for large $n$.

Below we review approximation techniques which have been proposed in the literature~\cite{han2012,chen2019,tsitsulin2018kdd}.
We empirically evaluate their approximation performance in Section~\ref{ssec:approx-acc}.

\subsection{Taylor Expansion}\label{ssec:taylor}
A natural impulse for dealing with complex matrix functions is to approximate the function with first few terms of its Taylor expansion.
Even though it is known that Taylor expansion provides an unreliable approximation of matrix functions~\cite{moler2003}, 
both NetLSD and VNGE rely on this approximation~\cite{tsitsulin2018kdd,chen2019,han2012,minello2018}, as the first two Taylor terms can be computed in $\bigO(m)$.

NetLSD's expansion depends on the parameter $t$, and its approximation is reasonable for only small $t$ values~\cite{tsitsulin2018kdd}:
\begin{equation}\label{eq:hks-taylor}
h_t = \sum_{i}{e^{-t\lambda_i}} = \sum_{k=0}^{\infty}{\frac{\tr({(-t\mLn)}^k)}{k!}}
    \approx n - t\, \tr(\mLn) + \frac{t^2}{2}\tr\mLn^2.
\end{equation}
The expansion used in VNGE is slightly different~\cite{chen2019,han2012,minello2018}:
\begin{equation}\label{eq:vnge-taylor}
\mHe = \sum_{i}{\lambda_i\ln\lambda_i} \approx 1 - \frac{1}{\tr(\mL)^2}(\tr(\mL) + 2\tr(\mL^2)).
\end{equation}
\noindent These first two terms are easily computed, even for very large graphs, as $\tr(\mathcal{L})=n$ and \(\tr\left(\mathcal{L}^2\right) = \sum_{ij}{{\mathcal{L}_{ij}}^2}\) since \(\mathcal{L}\) is self-adjoint, and the error rate of the Taylor expansion of the matrix exponential depends on the largest eigenvalue of the matrix~\cite{moler2003}.

Chen~et~al.~\cite{chen2019} introduce two approximation algorithms for VNGE based on a two-term Taylor expansion, FINGER-$\overline{\mHe}$ and FINGER-$\widehat{\mHe}$:
\begin{align*}\label{eq:vnge-finger}
\mathcal{Q} &=  1 - \frac{1}{\tr(\mL)^2}(\tr(\mD)^2 + 2\tr(\mL^2)) & \\
\mathrm{FINGER-}\overline{\mHe} &= -Q \ln\left(\frac{2\max{\mD}}{\tr(\mL)^2}\right), \qquad \mathrm{FINGER-}\widehat{\mHe} = -Q \ln\left(\lambda_{\max{}}\!\right)
\end{align*}

\subsection{Spectral Interpolation}

We conclude by noting that the Taylor expansion is useful on very large graphs, on which computing any part of the spectrum is prohibitive.
For manageable graph sizes, NetLSD adopts a more accurate strategy based on approximating the eigenvalue growth rate, adapted from~\cite{vaxman2010}.
It takes $\bigO(km+k^2n)$ to compute $k$ extremal eigenvalues of a graph~\cite{golub2009}, thus it is possible to compute $k$ eigenvalues on both ends of the spectrum, and interpolate a \emph{linear growth} of the interior eigenvalues.

While Tsitsulin~et~al.~\cite{tsitsulin2018kdd} do not provide approximation guarantees of their method,
it is easy to see that the worst-case scenario is the graph with exactly $k$ isolated nodes and a fully connected component having $n-k$ nodes, meaning $\lambda_{:k} = 0$ and $\lambda_{k:} = 2$.
Then, absolute error in the approximation of $h_t$ becomes $\| n - 2k - \sum_{i=0}^{n-2k} \frac{2(i-k)}{n-2k} \|$.
This bound is very loose; we further verify that the approximation accuracy of the linear interpolation strategy is poor in the Section~\ref{ssec:approx-acc} and that it does not scale to very large graphs in the Section~\ref{ssec:scalability}. %
\section{Stochastic Lanczos Quadrature}\label{sec:thiswork}
As noted above, the main approximation techniques that have been proposed for VNGE and NetLSD have limited guarantees on their approximation quality,
and these weak guarantees have not been fully explored in the literature.
In this section, we address these deficiencies and propose our method for improved approximation of spectral distances between graphs.

\subsection{Trace Function Estimation}\label{ssec:computation}
Setting aside computational infeasibility of the na{\"i}ve eigenvalues calculation, loose Taylor expansion error bounds and linear interpolation heuristics, we attain theoretically guaranteed accuracy and speed by means of Stochastic Lanczos Quadrature (SLQ)~\cite{ubaru2017}.

In trace estimation problems for large and implicit matrices, the standard choice is a Hutchinson estimator~\cite{hutchinson1989, adams2018}, which we apply (we denote both Laplacians as $\mL$ in this section for brevity) in our setting for the trace of matrix exponential $f(\mL) = \exp(-t \mL)$ or the matrix logarithm $f(\mL) = - \mL \log \mL$:
\begin{equation}\label{eq:trace}
 	\tr(f(\mL)) = \mathbb{E}_{p(\vV)}(\vV^\top  f(\mL) \vV ) \approx \frac{n}{n_v} \sum_{i=1}^{n_v} \vV_i^\top f(\mL) \vV_i,
 \end{equation}
\noindent where $\vV_i$ are $n_v$ random vectors drawn from a distribution $p(\vV)$ with zero mean and unit variance.
Practical choices for $p(\vV)$ include Rademacher or standard normal distributions, with the difference being in the variance or number of random vectors~\cite{avron2011}.

To approximate the bilinear form $\vV_i^\top f(\mL) \vV_i$ in Eq.~\eqref{eq:trace} with a symmetric real-valued matrix $\mL$, we apply the Lanczos Quadrature~\cite{golub2009}, which uses Lanczos algorithm to provide orthonormal polynomials for the Gauss quadrature. In other words, we first take an eigendecomposition $\mL = \mPhi\mLambda\mPhi^\top$, then cast the outcome to a Riemann-Stieltjes integral and finally apply the $m$-point Gauss quadrature rule:
\begin{align*}
\vV_i^\top f(\mL) \vV_i &= \vV_i^\top  \Phi f(\mLambda)\Phi^\top \vV_i = \sum\limits_{j=1}^n f(\lambda_j)\mu_j^2 \\ &=
\int_a^b f(t) d\mu(t) \approx \sum\limits_{k=1}^m \omega_k f(\theta_k),
\end{align*}
\noindent where $\mu_j = [\Phi^\top \vV_i]_j$ and $\mu(t)$ is a piecewise constant measure function:
$$\mu(t) = \begin{cases}
			0, &\text{if } t<a = \lambda_n \\
			\sum_{j=1}^i \mu_j^2, &\text{if } \lambda_i\leq t < \lambda_{i-1} \\
			\sum_{j=1}^n \mu_j^2, &\text{if } b=\lambda_1\leq t
		   \end{cases}
$$
\noindent and $\theta_k, \omega_k$ are the nodes and weights of the quadrature.
We obtain the pairs of $\omega_k, \theta_k$ with the $s$-step Lanczos algorithm~\cite{golub2009}, which we describe below succinctly.

The $s$-step Lanczos algorithm computes an orthonormal basis for the Krylov subspace $\mathcal{K}$ spanning vectors $\{\vQ_0, \mL \vQ_0, \dots, \mL^{s-1}\vQ_0\}$, with the symmetric matrix $\mL$ and an arbitrary \emph{starting unit-vector} $\vQ_0$.
The output of the algorithm is an $n \times s$ matrix $\mQ = [\vQ_0, \vQ_1, \dots, \vQ_{s-1}]$ with orthonormal columns and an $s \times s$ tridiagonal symmetric matrix $\mT$, such that $\mQ^\top \mL \mQ = \mT$, notice that due to this relation each $\vQ_i$ vector is given as a polynomial in $\mL$ applied to the initial vector $\vQ_0$: $\vQ_i = p_i(\mL)\vQ_0$. Since $\mT$ is a tridiagonal matrix, the three-term recurrence relation exists between the consequent polynomials $p_i$. We can now use the Gauss rule with points equal to the eigenvalues of $\mT$, $\lambda_k$, and weights set to the squared first components of its normalized eigenvectors, $\tau_k^2$, respectively (see \cite{golub1969calculation, wilf1962mathematics, golub2009}). 
Now, setting $\vQ_0 = \vV_i$, the estimate for the quadratic form becomes:
\begin{align}\label{eq:quadratic}
	&\vV_i^\top f(\mL) \vV_i \approx \sum\limits_{k=0}^{s-1} \tau_k^2 f(\lambda_k),  \\
	&\tau_k = \mU_{0,k} = \vE_1^\top \vU_k, \quad
	\lambda_k = \mLambda_{k,k} \quad \mT = \mU \mLambda \mU^\top
\end{align}
Applying~\eqref{eq:quadratic} over $n_v$ random vectors in the Hutchinson trace estimator~\eqref{eq:trace} yields the SLQ estimate:
\begin{equation}\label{eq:slq}
	\tr(f(\mL)) \approx \frac{n}{n_v} \sum_{i=0}^{n_v-1} \left(\sum\limits_{k=0}^{s-1} \left(\tau_k^i\right)^2 f\!\left(\lambda_k^i\right)\right) = \Gamma.
\end{equation}
For the matrix exponential used in NetLSD, \cite{tsitsulin2019intrinsic} suggests that we do not need many Lanczos steps $s$ to achieve error $\epsilon$,
\begin{subnumcases}{\label{eq:eps2} \epsilon \leq }
{20 e^{-s^2/(2.5 t)}}, & $\sqrt{2t} \leq s \leq t$ \label{eq:eps2a}\\
{40t^{-1} e^{-0.5t} \Big( \frac{0.5et}{s}\Big)^s}, & $ s \geq t$ \label{eq:eps2b}
\end{subnumcases}

Another source of error lies in the Monte Carlo estimation of the trace as a mean of the quadratic forms $\vV^\top f(\mL) \vV$. To reduce the variance of the estimate, we apply the 2-term polynomial variance reduction technique.

Although matrix-vector product approximation bounds for the matrix exponential have been well studied~\cite{hochbruck1997}, analogous error estimates for the von Neumann entropy remain an open problem in numerical linear algebra.
We summarize the overall \thiswork method in Algorithm \ref{alg:ourwork} for both LSD and VNGE.

\begin{algorithm}[!t]
    \begin{algorithmic}[1]
    \Function{\thiswork\_LSD}{$G, s, n_v$} %
        \State{$\mL \gets \mathtt{Laplacian}(G)$}
        \State{$\mathtt{descriptor} \gets \mathtt{slq}(\mL, s, n_v, \exp(\mathtt{x})$)}
        \State{\Return{$\mathtt{descriptor}$}}
    \EndFunction{}
    
    \Function{\thiswork\_VNGE}{$G, s, n_v$} %
        \State{$\mP \gets \mathtt{DensityMatrix}(G)$}
        \State{$\mathtt{descriptor} \gets \mathtt{slq}(\mP, s, n_v, \mathtt{x}\ln(\mathtt{x})$)}
        \State{\Return{$\mathtt{descriptor}$}}
    \EndFunction{}

    \Function{$\mathtt{slq}$}{$\mL, s, n_v, \mathtt{fun}$}
        \State{$\mT = \mathtt{lanczos}(\mL, s, n_v)$} \Comment{} $\mT \in \mathrm{R}^{n_v \times m \times m}$%
        \State{$\mLambda, \mU \gets \mathtt{eigh}(\mT)$} \Comment{} $\mathtt{eigendecomposition}(\mT)$
        \State{\Return{$\frac{1}{n_v} \sum\limits_{i}^{n_v}{\Big(\sum\limits_{k}^{s}(\mathtt{fun}(\lambda^i_{k}) [\vU^i_{k,0}]^2)\Big)}$}}
    \EndFunction{}
    \end{algorithmic}
    \caption{\thiswork algorithm}\label{alg:ourwork}
\end{algorithm} %

\section{Experiments}\label{sec:experiments}

We evaluate \thiswork{} against all approximation methods proposed in~\cite{chen2019,tsitsulin2018kdd}, in addition to the exact computation of the spectrum (where allowed by the graph size).
We perform our experiments on the Google Cloud's \texttt{c2-standard-60} virtual machine with 60 virtual cores and 240GB RAM, averaging 10 times for all experiments unless stated otherwise.
We use LAPACK~\cite{anderson1999} as the linear algebra library of choice. We open-source the implementation\footnote{\url{github.com/google-research/google-research/tree/master/graph_embedding/slaq}}.

\mpara{Parameter settings.} Unless otherwise mentioned, we evaluate \thiswork using $n_v=100$ starting vectors and $s=10$ Lanczos iterations.
We provide an additional experimental investigation into parameter settings of \thiswork{} in Section \ref{ssec:paramter-sensitivity}.
For the linear approximation of~\cite{tsitsulin2018kdd}, we use the default ($k=300$) eigenvalues from each end of the spectrum, following the notation of the original paper.
Taylor series-based approximation techniques do not depend on any additional parameters.

\mpara{Datasets.}
We use four types of graph collections to measure efficiency and effectiveness of \thiswork.
First, we consider the accuracy of the method compared to other approximation techniques on the two subsets of graphs: synthetically generated Erdos-Renyi graphs and 73 graphs from the Network Repository\footnote{\url{networkrepository.com}}~\cite{rossi2015} with a number of nodes from 2500 up to 25000.
In total, we use 27 biological, 12 interaction, 10 technological networks, 5 small web graphs, and 19 uncategorized networks (mostly, optimization problem graphs).

We follow up with large graphs to test ability of \thiswork{} to efficiently compute descriptors of Web-scale graphs.
For that, we use five datasets\footnote{All but \dclueweb{} are from SNAP network collection, available at \url{snap.stanford.edu}}:
\begin{itemize}[leftmargin=0cm,itemindent=.4cm,labelwidth=\itemindent,labelsep=0cm,align=left]
\item \ddblp{}~\cite{yang2012} is a co-authorship network constructed from DBLP, a major online computer science bibliography resource. 
\item \dorkut{}~\cite{yang2012} was an online social network.
\item \dlivejournal{}~\cite{yang2012} is an online blogging community where users can form friendships with each other.
\item \dfriendster{}~\cite{yang2012} was an online social network.
\item \dclueweb{}~\cite{clarke2009,rossi2015} is a web crawl from 2009.
\end{itemize}

Next, we investigate benefits of using \thiswork{} on dynamic Wikipedia link datasets in 4 different languages:
Dutch (\texttt{nl}), Polish (\texttt{pl}), Italian (\texttt{it}), German (\texttt{de}).
We obtained datasets from~\cite{clarke2009}\footnote{We used preprocessed version from KONECT repository, available at \url{konect.uni-koblenz.de/networks/}} and generated $|T|$ snapshots for every month in the original dataset.

\begin{table}[!t]
\setlength{\tabcolsep}{3.5pt}
\centering{
\newcolumntype{R}{>{\raggedleft\arraybackslash}X}
\newcolumntype{C}{>{\centering\arraybackslash}X}
\begin{tabularx}{\linewidth}{p{1.7cm}CCCCR}
\multicolumn{1}{c}{} & \multicolumn{2}{c}{\textbf{Size}} & \multicolumn{2}{c}{\textbf{Statistics}} \\
\cmidrule(lr){2-3}\cmidrule(lr){4-5}
\emph{dataset} & \(|V|\) & \multicolumn{1}{c}{\(|E|\)} & \mbox{Avg.\ deg.} & Density \\
    \midrule
\small\ddblp{} & 317k & 1.05M & 6.62 & $2.08 \times 10^{-5}$ \\
\small\dorkut{} & 3.07M & 117.2M & 76.28 & $2.48 \times 10^{-5}$ \\
\small\dlivejournal{} & 4M & 34.7M & 17.35 & $4.34 \times 10^{-6}$ \\ 
\small\dfriendster{} & 65.6M & 1.8B & 55.06 &  $8.39 \times 10^{-7}$ \\
\small\dclueweb{} & 4.8B & 7.81B & 3.27 & \mbox{$6.83 \times 10^{-10}$} \\
\bottomrule
\end{tabularx}}
\caption{Characteristics of large graphs used in this work: number of vertices \(|V|\), number of edges \(|E|\); average node degree; density defined as \(|E|/\binom{|V|}{2}\).}\label{tbl:datasets-large}
\vspace{-3mm}
\end{table} %

\begin{table}[!t]
\setlength{\tabcolsep}{3.5pt}
\centering{
\newcolumntype{R}{>{\raggedleft\arraybackslash}X}
\newcolumntype{C}{>{\centering\arraybackslash}X}
\begin{tabularx}{\linewidth}{p{1.7cm}CCCCR}
\multicolumn{1}{c}{} & \multicolumn{2}{c}{\textbf{Size}} & \multicolumn{2}{c}{\textbf{Temporal statistics}} \\
\cmidrule(lr){2-3}\cmidrule(lr){4-5}
\emph{dataset} & \(|V|\) & \multicolumn{1}{c}{\(|E|\)} & $|T|$ & $|\mathcal{E}|/|T|$ \\
    \midrule
\small\dwikim{nl} & 1M & 20M & 95 & 148337 \\
\small\dwikim{pl} & 1M & 25M & 95 & 182959 \\
\small\dwikim{it} & 1.2M & 35M & 95 & 250633 \\
\small\dwikim{de} & 2.1M & 86M & 95 & 553257 \\
\bottomrule
\end{tabularx}}
\caption{Characteristics of dynamic graphs: total number of vertices \(|V|\), total number of edges \(|E|\); number of timestamps $|T|$; average incoming edges per timestamp $|\mathcal{E}|/|T|$.}\label{tbl:datasets-dynamic}
\vspace{-3mm}
\end{table}

\begin{table}[!t]
\begin{center}
{
\newcolumntype{C}{>{\centering\arraybackslash}X}
\begin{tabularx}{\columnwidth}{p{1.75cm}CCCCC}
& & & \multicolumn{3}{c}{\textbf{Vertices $|V|$}}\\
\cmidrule(lr){4-6}
\emph{dataset} & $|G|$ & $|Y|$ & Min. & Avg. & Max. \\
\midrule
\ddd{} & 1178 & 2 & 30 & 284.32 & 5748 \\
\dcollab{} & 5000 & 3 & 32 & 74.49 & 492 \\
\mbox{\dredditm{}} & 4999 & 5 & 22 & 508.52 & 3648 \\
\mbox{\dredditmb{}} & 22939 & 11 & 2 & 391.41 & 3782 \\
\bottomrule
\end{tabularx}
}
\end{center}
\caption{Properties of the graph classification datasets used: number of graphs $|G|$; number of labels $|Y|$; minimum, average, and maximum number of nodes in graph collection.}\label{tbl:datasets-graphclassification}
\vspace{-3mm}
\end{table} 
Last, we verify that \thiswork{}'s improvements in approximation performance enhance downstream task performance.
We use three social network datasets and one from the field of bioinformatics\footnote{We obtained them at the graph kernel benchmark collection, available at \url{ls11-www.cs.tu-dortmund.de/staff/morris/graphkerneldatasets}}:
\begin{itemize}[leftmargin=0cm,itemindent=.4cm,labelwidth=\itemindent,labelsep=0cm,align=left]
\item \ddd{}~\cite{dobson2003,shervashidze2011,kersting2016} is a dataset of protein structures.
Each protein is represented by a graph of amino acids that are connected by an edge if they are less than 6 {\r{A}}ngstroms apart.
The prediction task is to classify the protein structures into enzymes and non-enzymes.
\item \dcollab{}~\cite{yanardag2015,leskovec2005,kersting2016} is a collection of collaboration ego-networks of different researchers derived in~\cite{yanardag2015} from three datasets introduced in~\cite{leskovec2005}.
The task is to determine whether the ego-collaboration graph of a researcher belongs to High Energy, Condensed Matter or Astrophysics field.
\item \dredditm{} and \dredditmb{}~\cite{yanardag2015,kersting2016} are two datasets derived from Reddit, an online aggregation and discussion website.
Discussions on Reddit are organized into different subcommunities; the task is to determine the community given the structure of the discussion graph.
\end{itemize}

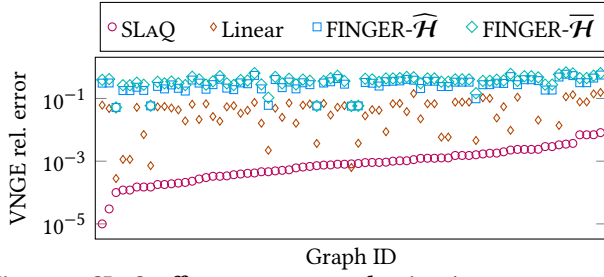
\begin{figure}[!t]
\centering
\begin{tikzpicture}
\begin{axis}[
    only marks,
	ylabel={VNGE rel.\ error},
	xlabel={Graph ID},
	legend columns=4,
	legend style={at={(1,1.03)},anchor=south east},
	ymode=log,
	width=0.99\columnwidth,
	height=4cm,
	xmin=-1,
	xmax=73,
	xtick style={draw=none},
	xtick=\empty,
]
\addplot[mark=o,mark options={color=cycle2}, mark size=1.5pt] table [x expr=\coordindex, y=slqvnge] {data/vnge-error.dat};\addlegendentry{\thiswork\vphantom{$\widehat{\mHe}$}\quad\quad};
\addplot[mark=diamond,mark options={color=cycle7}, mark size=1.5pt] table [x expr=\coordindex, y=inter] {data/vnge-error.dat};\addlegendentry{Linear\vphantom{$\widehat{\mHe}$}\quad\quad};
\addplot[mark=square,mark options={color=cycle3}, mark size=1.5pt] table [x expr=\coordindex, y=chenlambda] {data/vnge-error.dat};\addlegendentry{FINGER-$\widehat{\mHe}$\quad\quad};
\addplot[mark=square,mark options={color=cycle5, rotate=45}, mark size=1.5pt] table [x expr=\coordindex, y=chensimple] {data/vnge-error.dat};\addlegendentry{FINGER-$\overline{\mHe}$};
\end{axis}
\end{tikzpicture}
\vspace*{-5mm}
\caption{\thiswork{} offers over $\mathbf{200\times}$ reduction in average error for VNGE over techniques proposed in~\cite{chen2019} and over $\mathbf{30\times}$ improvement over the linear approximation  from~\cite{tsitsulin2018kdd}.}\label{fig:vnge-approximation-nr}
\vspace{-3mm}
\end{figure}
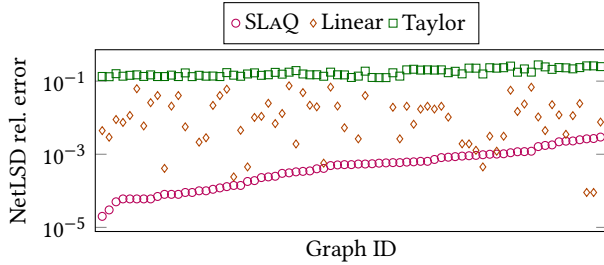
\begin{figure}
\centering
\begin{tikzpicture}
\begin{axis}[
    only marks,
	ylabel={NetLSD rel.\ error},
	xlabel={Graph ID},
	legend columns=3 ,
	legend style={at={(0.5,1.03)},anchor=south},
	ymode=log,
	width=0.99\columnwidth,
	height=4cm,
	xmin=-1,
	xmax=73,
	xtick style={draw=none},
	xtick=\empty,
]
\addplot[mark=o,mark options={color=cycle2}, mark size=1.5pt] table [x expr=\coordindex, y=slqlsd] {data/lsd-error.dat};\addlegendentry{\thiswork\quad};
\addplot[mark=diamond,mark options={color=cycle7}, mark size=1.5pt] table [x expr=\coordindex, y=inter] {data/lsd-error.dat};\addlegendentry{Linear\quad};
\addplot[mark=square,mark options={color=cycle4}, mark size=1.5pt] table [x expr=\coordindex, y=taylor] {data/lsd-error.dat};\addlegendentry{Taylor};\end{axis}
\end{tikzpicture}
\vspace*{-5mm}
\caption{\thiswork{} offers $\mathbf{22\:\!\times}$ reduction in average error for NetLSD over~\cite{tsitsulin2018kdd} and $\mathbf{250\:\!\times}$ over Taylor expansion.}\label{fig:lsd-approximation-nr}
\vspace{-3mm}
\end{figure}
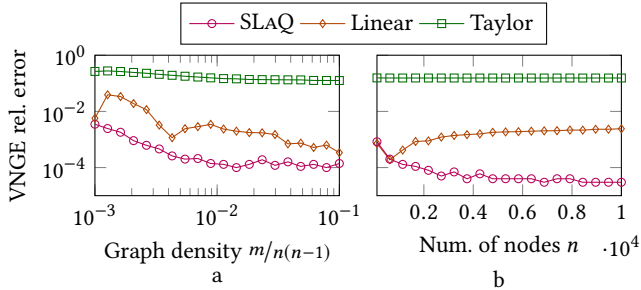
\begin{figure}[!t]
\centering
\begin{tikzpicture}
\begin{groupplot}[group style={
                      group name=myplot,
                      group size= 2 by 1, horizontal sep=0.5cm},ymin=0.00001,ymax=1,height=3.45cm,width=0.57\columnwidth,ymode=log,title style={at={(0.5,0)},anchor=north,yshift=-11mm}]
\nextgroupplot[
	ylabel={VNGE rel.\ error},
	xlabel={Graph density $\nicefrac{m}{n(n-1)}$},
	xmin=0.001,
	xmax=0.1,
	xmode=log,
 	title = {a},
	legend columns=3,
	legend style={at={(1.85,1.075)},anchor=south east},
    legend entries={\thiswork{}, Linear, Taylor},
]
\addplot[color=cycle2,mark=o, mark size=1.5pt] table [x=p, y=slq] {data/er-density.dat};
\addplot[color=cycle7,mark=diamond, mark size=1.5pt] table [x=p, y=inter] {data/er-density.dat};
\addplot[color=cycle4,mark=square, mark size=1.5pt] table [x=p, y=taylor] {data/er-density.dat};
\nextgroupplot[
	xlabel={Num.\ of nodes $n$},
	xmin=100,
	xmax=10000,
 	yticklabels={,,},
 	title = {b},
    every x tick scale label/.style={at={(xticklabel cs:1)},anchor=north},
]
\addplot[color=cycle4,mark=square, mark size=1.5pt] table [x=n, y=taylor] {data/er-nodes.dat};
\addplot[color=cycle7,mark=diamond, mark size=1.5pt] table [x=n, y=inter] {data/er-nodes.dat};
\addplot[color=cycle2,mark=o, mark size=1.5pt] table [x=n, y=slq] {data/er-nodes.dat};
\end{groupplot}
\end{tikzpicture}
\vspace*{-7mm} %
\caption{Number of nodes and edges of random Erd{\H o}s-R{\'e}nyi graphs does not affect \thiswork{}'s approximation accuracy.}\label{fig:approximation-accuracy-erdosrenyi}
\vspace{-3mm}
\end{figure}

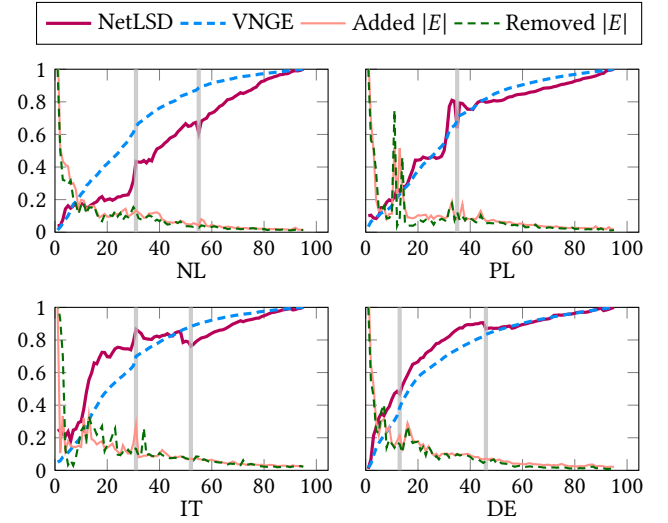
\begin{figure}[!t]
\centering
\begin{tikzpicture}
\begin{groupplot}[group style={
                      group name=myplot,
                      group size= 2 by 2, horizontal sep=0.5cm},height=3.75cm,width=0.615 \columnwidth,ymin=0,ymax=1,title style={at={(0.5,0)},anchor=north,yshift=-5mm}]
\nextgroupplot[
	xmin=0,
 	title = {NL},
 	legend columns=4,
	legend style={at={(2.15,1.15)},anchor=south east},
    legend entries={NetLSD, VNGE, Added $|E|$, Removed $|E|$},
]
\addplot[color=cycle2, very thick] table [x=t, y=lsd] {data/langs/nl.dat};\label{line:netlsd};
\draw [color=cyclegray, ultra thick, draw opacity=0.75] (31,0) -- (31,1);
\draw [color=cyclegray, ultra thick, draw opacity=0.75] (55,0) -- (55,1);
\addplot[color=cycle3, very thick, densely dashed] table [x=t, y=vnge] {data/langs/nl.dat};\label{line:vnge};
\addplot[color=cycle8, thick] table [x=t, y=add] {data/langs/nl.dat};
\addplot[color=cycle4, thick, densely dashed] table [x=t, y=del] {data/langs/nl.dat};
\nextgroupplot[
	xmin=0,
	yticklabels={,,},
 	title = {PL},
]
\addplot[color=cycle2, very thick] table [x=t, y=lsd] {data/langs/pl.dat};
\draw [color=cyclegray, ultra thick, draw opacity=0.75] (35,0) -- (35,1);
\addplot[color=cycle3, very thick, densely dashed] table [x=t, y=vnge] {data/langs/pl.dat};
\addplot[color=cycle8, thick] table [x=t, y=add] {data/langs/pl.dat};
\addplot[color=cycle4, thick, densely dashed] table [x=t, y=del] {data/langs/pl.dat};
\nextgroupplot[
	xmin=0,
 	title = {IT},
]
\addplot[color=cycle2, very thick] table [x=t, y=lsd] {data/langs/it.dat};
\draw [color=cyclegray, ultra thick, draw opacity=0.75] (31,0) -- (31,1);
\draw [color=cyclegray, ultra thick, draw opacity=0.75] (52,0) -- (52,1);
\addplot[color=cycle3, very thick, densely dashed] table [x=t, y=vnge] {data/langs/it.dat};
\addplot[color=cycle8, thick] table [x=t, y=add] {data/langs/it.dat};
\addplot[color=cycle4, thick, densely dashed] table [x=t, y=del] {data/langs/it.dat};
\nextgroupplot[
	xmin=0,
	yticklabels={,,},
 	title = {DE},
]
\addplot[color=cycle2, very thick] table [x=t, y=lsd] {data/langs/de.dat};
\draw [color=cyclegray, ultra thick, draw opacity=0.75] (13,0) -- (13,1);
\draw [color=cyclegray, ultra thick, draw opacity=0.75] (46,0) -- (46,1);
\addplot[color=cycle3, very thick, densely dashed] table [x=t, y=vnge] {data/langs/de.dat};
\addplot[color=cycle8, thick] table [x=t, y=add] {data/langs/de.dat};
\addplot[color=cycle4, thick, densely dashed] table [x=t, y=del] {data/langs/de.dat};
\end{groupplot}
\end{tikzpicture}
\vspace*{-5mm} %
\caption{\thiswork{} approximation of~\ref{line:netlsd} NetLSD and~\ref{line:vnge} VNGE for Wikipedia graphs across time. Changes that are not explained by local edge differences highlighted in gray.}\label{fig:spectral-consistency}
\vspace{-3mm}
\end{figure}

\subsection{Approximation Accuracy}\label{ssec:approx-acc}
We proceed with evaluation of \thiswork capacity to approximate matrix functions for graph comparison.
We compute full spectrum of 73 small graphs and true values of NetLSD and VNGE and report the relative $l_2$ approximation error with respect to the true graph descriptor.
Figure~\ref{fig:vnge-approximation-nr} demonstrates that \thiswork{} offers over $200\:\!\times$ reduction in the average relative error for VNGE over FINGER techniques~\cite{chen2019} and over $30\:\!\times$ improvements over the linear interpolation technique from~\cite{tsitsulin2018kdd}.
Figure~\ref{fig:lsd-approximation-nr} shows that \thiswork{} offers $250\:\!\times$ reduction over the Taylor expansion and over $22\:\!\times$ improvement over the linear interpolation~\cite{tsitsulin2018kdd}.

We also compare how the approximation accuracy changes for NetLSD on the random Erd{\H o}s-R{\'e}nyi graphs.
We generate random graphs of size $1000$ with varying graph density (number of expected edges) $p$ and random graphs of size $100-10000$ with the average degree $\nicefrac{m}{n}=10$.
We report the results in the Figure~\ref{fig:approximation-accuracy-erdosrenyi}.
We observe that \thiswork{}'s approximation accuracy is stable across the graph size both in terms of number of the nodes and graph density.

\subsection{Benefits of Non-local Approximation}
\label{ssec:temporal-consistency}
We verify that our method has a global view of the graph, i.e.\ is not dominated by only local information.
In order to do that, we compute graph descriptors (NetLSD and VNGE) for monthly snapshots of dynamic Wikipedia link datasets from January 2003 to December 2010 (a total of $|T| = 96$ snapshots) and report their change as well as the number of edges added/removed each month.

We plot the proportion of cumulative edge additions/deletions and distances between descriptor pairs of snapshots $(0, i)$, where $i \in 1, \dots, |T|$.
Figure~\ref{fig:spectral-consistency} reports the distance values for each language as well as the relative number of incoming and outgoing edges per snapshot.
We mark exmaples of anomalous spikes in NetLSD and VNGE that can not be explained simply by the edge additions and deletions.
In these cases, simple approximations like 2-term Taylor expansion would fail to capture such changes.
\begin{table*}[!th]
\begin{center}
{
\setlength{\tabcolsep}{0pt}
\renewcommand{\aboverulesep}{0pt}
\renewcommand{\belowrulesep}{0pt}
\newcolumntype{C}{>{\centering\arraybackslash}X}
\begin{tabularx}{\textwidth}{p{1.75cm}CCCCCCCCC}
& \multicolumn{5}{c}{\textbf{VNGE}} & \multicolumn{4}{c}{\textbf{NetLSD}} \\
\cmidrule(lr){2-6}\cmidrule(lr){7-10}
\emph{dataset} & \mbox{FINGER-$\overline{\mHe}$} & \mbox{FINGER-$\widehat{\mHe}$} & Linear & \thiswork{} & \textbf{Exact} & Taylor & Linear & \thiswork{} & \textbf{Exact} \\
\midrule
\ddd{} & 63.01 & \win 66.38 & \win 68.13 & 65.53 & \textbf{66.40} & \win 67.01 & \win 67.98 & 66.77 & \textbf{67.24} \\
\dcollab{} & \win 64.95 & \win 65.10 & 55.90 & 49.04 & \textbf{58.03} & 61.81 & \win 65.17 & 58.76 & \textbf{63.48} \\
\mbox{\dredditm{}} & 30.87 & 29.85 & \win 31.31 & \win 31.77 & \textbf{31.43} & 33.67 & 32.01 & \win 35.48 & \textbf{35.63} \\
\mbox{\dredditmb{}} & \win 16.53 & 16.20 & \win 17.18 & \win 17.04 & \textbf{16.79} & 22.67 & 21.30 & \win 25.31 & \textbf{25.52} \\
\bottomrule
\end{tabularx}
}
\end{center}
\caption{1-Nearest neighbor graph classification performance on 4 datasets with VNGE and NetLSD. Exact computation results are in bold. Approximations that are close to or better than the exact metric computation are highlighted \colorbox{cycle4!25}{in green.} }\label{tbl:exp-graphclassification}
\vspace{-8mm}
\end{table*} %
\begin{figure}
\centering
\begin{tikzpicture}
\begin{groupplot}[group style={
                      group name=myplot,
                      group size= 2 by 1, horizontal sep=1.35cm},height=4cm,width=0.5\columnwidth,ymode=log,title style={at={(0.5,0)},anchor=north,yshift=-11mm}]
\nextgroupplot[
	ylabel={NetLSD rel.\ error},
	xlabel={Number of random vectors $n_v$},
	xmin=10,
	xmax=5000,
	xmode=log,
	ymin=0.0005,ymax=0.002,
 	title = {a},
]
\addplot[color=cycle2,mark=*, mark size=1.5pt] table [x=nv, y=error] {data/param-sensitivity-nv.tex};

\nextgroupplot[
	xlabel={Number of Lanczos steps $s$},
	xmin=0,
	xmax=100,
	xmode=log,
    ymin=0.0001,ymax=0.01,
 	title = {b},
]
\addplot[color=cycle2,mark=*, mark size=1.5pt] table [x=m, y=error] {data/param-sensitivity-m.tex};
\end{groupplot}
\end{tikzpicture}
\vspace*{-7mm} %
\caption{Parameter sensitivity of \thiswork{} in terms of approximating NetLSD with (a) different number of starting vectors $n_v$ and (b) different number of Lanczos steps $s$. Error averaged across 73 graphs from the Network Repository.}\label{fig:parameter-sensitibity}
\end{figure}
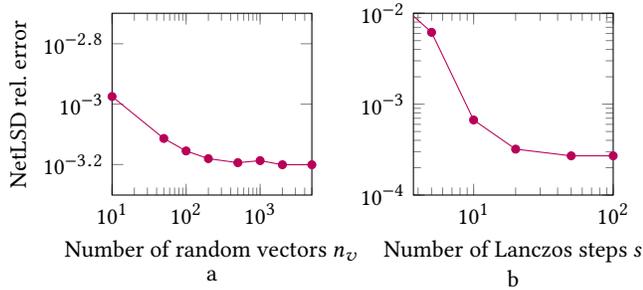 %

\begin{table}[!h]
\setlength{\tabcolsep}{3.5pt}
\centering{
\newcolumntype{R}{>{\raggedleft\arraybackslash}X}
\newcolumntype{C}{>{\centering\arraybackslash}X}
\newcolumntype{S}{>{\hsize=.5\hsize}C}
\begin{tabularx}{\linewidth}{p{1cm}CCCC}
\emph{dataset} & \mbox{FINGER-$\overline{\mHe}$} & \mbox{FINGER-$\widehat{\mHe}$} & Linear & \thiswork{} \\
    \midrule
\small\ddblp{} & 0.06 & 0.65 & 394 & 28.4 \\ 
\mbox{\small\dorkut{}} & 1.68 & 70 & 8863 & 899 \\ 
\small\dlivejournal{} & 0.97 & 15.6 & 4727 & 476 \\ 
\small\dfriendster{} & 1.67 & 71 & OOM & 900 \\ 
\small\dclueweb{} & 902 & OOM & OOM & 3447 \\ 
\bottomrule
\end{tabularx}}
\caption{Running time (in seconds) of different approximation techniques and \thiswork{} for VNGE on large graphs.}\label{tbl:exp-runtime}
\vspace{-7mm}
\end{table} 
\subsection{Graph Classification Performance}
\label{ssec:graph-class-perf}
We test our method in the supervised downstream task, by classify graphs in binary and multi-class settings.
We compute NetLSD and VNGE descriptors for each of the graphs and use them as feature vectors in classification.
Since these graph classification datasets allow direct calculation of the descriptor (maximum number of nodes reported in the Table \ref{tbl:datasets-graphclassification} is 5748), we can analyze how approximation affects the downstream accuracy.

We use a non-parametric 1-Nearest Neighbor classification algorithm and repeat the classification using 80/20 training/testing split 1000 times to minimize the biases introduced by the random splitting and the learning algorithm.
We report the classification accuracy in the Table~\ref{tbl:exp-graphclassification}.

Surprisingly, on two datasets, \ddd{} and \dcollab{}, the classification accuracy is actually \emph{better} for the low-accuracy approximation.
We believe that this relates to the issues with these datasets pointed out by~\cite{schulz2019,cai2019}: simple local graph features achieve almost state-of-the-art performance~\cite{alrfou2019}.
However, for the Reddit datasets the improvement given by more accurate approximation is as expected due to the task being more sensitive to global structural information rather than simple node-level statistics.

\subsection{Parameter Sensitivity}\label{ssec:paramter-sensitivity}
We investigate the approximation accuracy of \thiswork{} with respect to its hyperparameters: number of random starting vectors $n_v$ and the number of Lanczos iterations $s$.
Recall that the error bounds in the Section~\ref{sec:thiswork} tells us that there are two sources of error in \thiswork{}: one of the Monte Carlo estimation of the quadratic form and one of the Lanczos process.
We measure the relative error ratio on the same 73 medium-sized graphs used in the Section~\ref{ssec:approx-acc} with respect to the number of random starting vectors $n_v$ and the number of Lanczos iterations $s$ and report the results in the Figure~\ref{fig:parameter-sensitibity}.

As expected, given enough stating random vectors \thiswork{} only needs few Lanczos iterations; the default setting of $s=10$ gives an average error of $6.7\times10^{-4}$.
As for the number of random vectors $n_v$, we do observe that increasing the number improves performance, but the improvement given by increasing $n_v$ is much slower.

\subsection{Scalability}\label{ssec:scalability}
We measure the runtime of all approximation techniques on huge graphs with millions of nodes and billions of edges and show that \thiswork{} is able to process very large graphs on a single machine within reasonable time while offering orders of magnitude better approximation, as measured in the Section~\ref{ssec:approx-acc}.
We only report the results for VNGE, as the results for NetLSD for Linear interpolation and \thiswork{} approximation are similar to VNGE counterparts, while Taylor approximation works in the same time as FINGER-$\overline{\mHe}$.

As FINGER-$\overline{\mHe}$ only sums the weights of graphs' edges, it serves as a baseline on how much time it takes to scan the edges of a graph.
A more useful comparison is FINGER-$\widehat{\mHe}$, as it reflects the time to compute a \emph{single} eigenvalue of a graph.
\thiswork{} approximates the \emph{whole} spectrum at the cost of increased time complexity, however, the largest dataset with almost 5 billion nodes is processed in less than an hour.

\vspace{-2mm}
\section{Conclusion}\label{sec:conclusion}

We propose \thiswork{}, an approximation technique for fast computation of spectral graph distances, VNGE and NetLSD, leveraging state-of-the-art linear algebra methods.
We show that faithful approximation of the graph distance is critical for good downstream task performance and those approximation methods previously introduced in the literature do not offer good approximation quality.
\thiswork{} improves approximation errors of such baseline solutions by at least an order of magnitude averaged across 73 real-world graphs.
As \thiswork{}'s computation is linear in the number of edges of graphs, scalability of our method is on par with approximation techniques introduced for VNGE and NetLSD.
To our knowledge, this is the first work that allows accurate comparison of billion-size graphs on a single machine in less than an hour. 
\newpage
\balance
\bibliographystyle{ACM-Reference-Format}
\bibliography{main.bib}
\end{document}